\UseRawInputEncoding 
\documentclass{resonance}
\usepackage{hyperref}

\begin{document}

\title{Starlight in Darkness:}
\secondTitle{The Birth of Stars}
\author{Priya Hasan}

\maketitle
\authorIntro{\includegraphics[width=2cm]{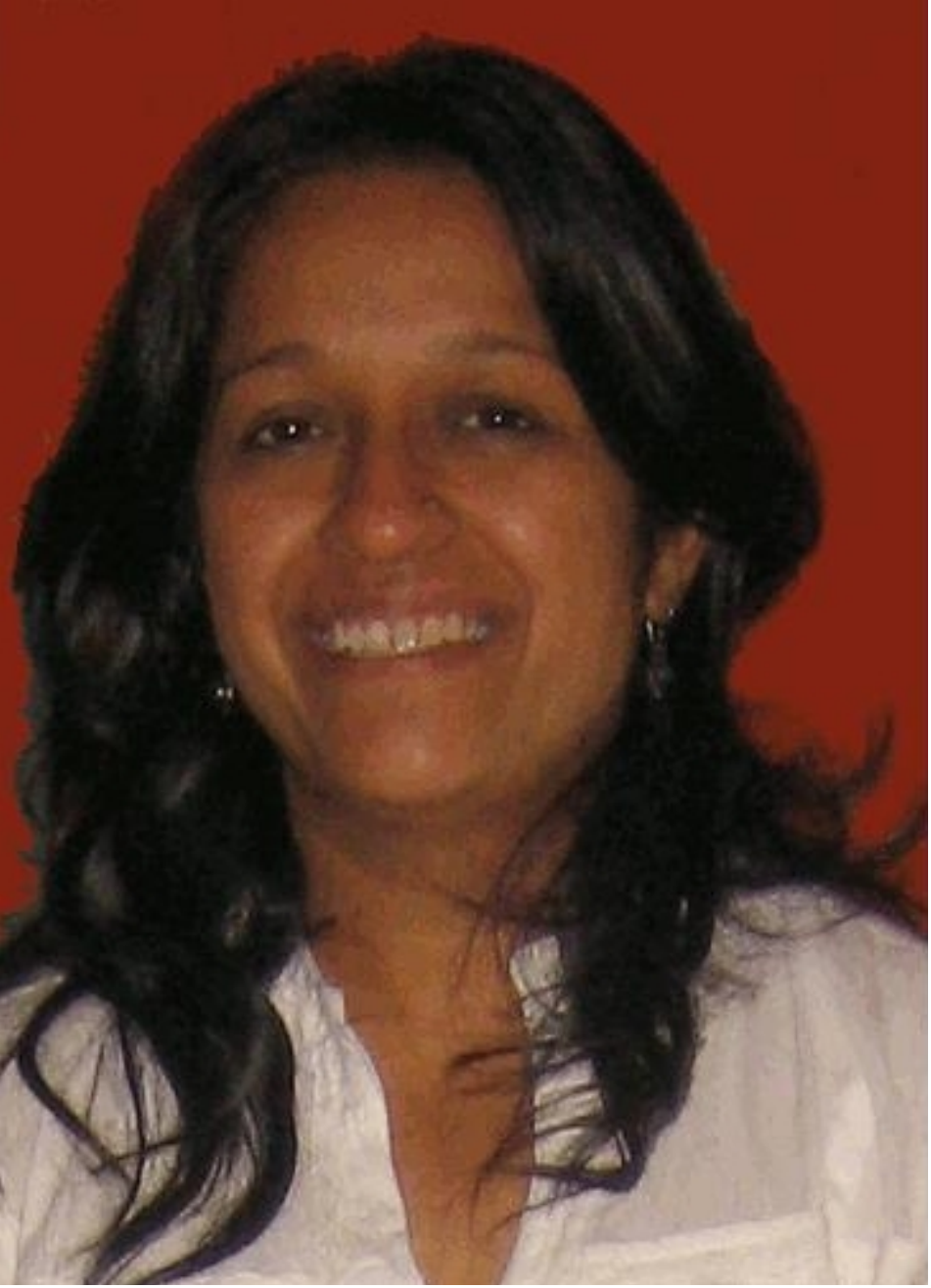}\\
Dr Priya Hasan is an Asst Professor in Physics at the Maulana Azad National Urdu University, Hyderabad. Her research interests are in observational astronomy, star formation, star clusters and galaxies.  She is actively involved in olympiads, public outreach and science popularization programs. She is a member of the Astronomical Society of India - Public  Outreach \& Education Committee (2016-2019)}
\begin{abstract}
This article will briefly review the theory of star formation (SF) and its development using observations. This is very relevant in the present context since planet formation appears to be a  byproduct of SF, and the whole question of life in the universe and its origin can be viewed with a new perspective.
\end{abstract}

\monthyear{March 2017}
\artNature{GENERAL  ARTICLE}

\begin{center}

\it{If the stars should appear one night in  a thousand years, how would men believe  and adore, and preserve for many generations  the remembrance of the city of God?
  \\
 Nightfall, Isaac Asimov}
 \end{center}

\section*{Introduction}
Humans have been bestowed with a gift, from the heavens: the gift of starlight. Imagine that we were on a planet, like Venus,  with an opaque atmosphere. Or like the people of Lagash in  `Nightfall',  a 1941 science-fiction novelette by Isaac Asimov. Their planet is lit by six suns and they wait for the year 2049 when eclipsing of all the suns would lead to  the night of darkness which would reveal stars that madden the people...

 We  live in a unique world spaced by day and night, reminders of light and darkness,  ...each with its own beauty.
\keywords{molecular clouds, Jean's criterion, protostars, protoplanetary discs, star clusters}  
The most enchanting component of our skies are stars. To our ancestors, the most  obvious explanation was that they were lanterns, or fireplaces in the skies. Then they noticed patterns, that they moved, in concentric circles, about a fixed point in approximately 24 hours. And hence followed the notion of  a shell of stars rotating around the earth, with the fixed points as the north and  south poles.

%

\begin{figure}[!t]
\caption{Structure of the Milky Way. Image credit: http://www.profjohn.com/courses/ast1004/milkyway/milkyway1.htm}
\label{gal}
\vskip -12pt
\centering
\includegraphics[width=10cm, height=5.5cm]{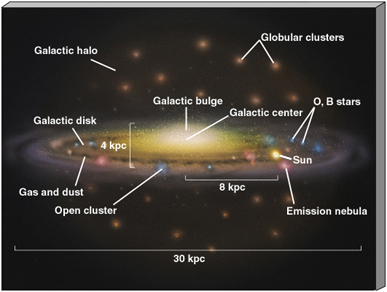}
\end{figure}

The stars of a galaxy distribute themselves into broadly three components, viz.: the disc, halo and bulge (Fig.\ref{gal}).  The halo is made up of an older population of stars that constitute globular clusters. Globular clusters are made up of low metallicity, dense aggregates of 50,000 to 100,000 stars, gravitationally bound, with orbits that are randomly distributed, which leads to their spherical shape. The stars are redder, older $\approx$ 10 Gyr\footnote {Common Units: 1 Myr = $10^6$ yr, 1 Gyr = $10^8$ yr, 1 parsec (pc) = $3 \times 10^{13}$ km, 1 light year (ly)= $10^{13}$ km, 1 $M_{\odot}$ = Mass of the Sun = 2 $\times  10^{30}$ kg} and we do not see any signs of SF taking place there. The disc component is made up of spiral arms where young stars are forming, even now, as it is gas rich and this is where we find open star clusters which are looser aggregates of stars with typical lifetimes of a few 100 Myr$^1$. The nuclear bulge contains the highest density of stars in the galaxy. 

\rightHighlight{The Milky Way is made up of three components: disc, halo, bulge. Active SF is taking place in the disc. Star clusters are found in a continuous range of densities of 0.01-100 stars pc$^{-3}$, with the older globular clusters in the halo and the younger associations and open star clusters in the gas rich discs$^1$.}

Although some hot young stars may be found in the nucleus, the primary population of stars there is similar to the old stars found in the halo. The core of the galaxy is highly obscured by dust and gas at visible wavelengths and can be observed  at other wavelengths. Our galaxy contains a very massive black hole at its center with a mass of $\approx 4.5\times 10^6$ M$_\odot^1$, which drives many of these processes near the core.

\begin{figure}[!t]
\caption{Typical density of associations, open clusters and globular clusters.(Image : 	
Moraux, Estelle; Lebreton, Y; Charbonnel, C, EAS Publications Series, Volume 80-81, Stellar Clusters: Benchmarks of Stellar Physics and Galactic Evolution)}
\label{clus}
\vskip -12pt
\centering
\includegraphics[width=9.5cm, height=3cm]{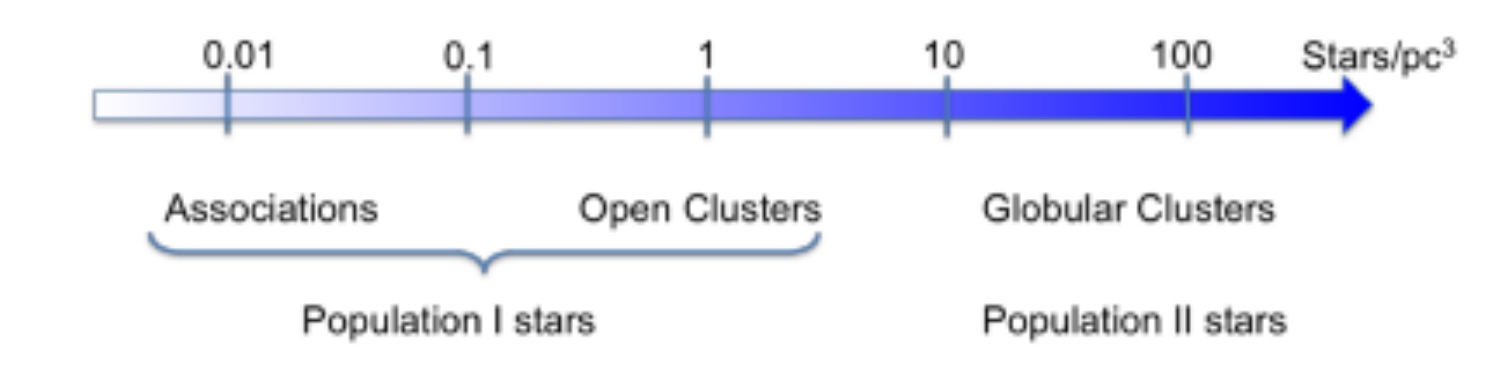}
\end{figure}

 Star clusters are nature's samples of stars formed from the same parent cloud and therefore of the same age, chemical composition and at the same distance, differing only in mass. The  Russell-Vogt theorem states  that mass and chemical composition are the two important parameters that decide a star's fate. Hence, star clusters are widely used as ideal samples to study stellar evolution as all other parameters are fixed and the mass of stars defines its' evolution. In the present times, they are also very useful in understanding star and planet formation as these are very closely linked processes, planet formation being a byproduct of SF.
\rightHighlight{Open clusters, associations, and moving groups are probably just different realisations of the SF process, differentiated due to the way in which we observe them, their environments and other factors.} 
There are also associations, which consist of recently formed stars, not bound gravitationally, at large separations of $\approx$ 100 pc, and expanding away from some common center, which presumably marks their birthplace. The motion of these stars can be traced back in time to support this conjecture. They are categorized as OB,  T and R Associations, based on the properties of their stars, which could vary. (Fig. \ref{clus}). If the remnants of a stellar association drift through the Milky Way as a  coherent set of stars, they are called a moving group or kinematic group.

\section{Voids in in the Milky Way: Molecular Clouds}

\begin{figure}[!t]
\caption{BVI image of molecular cloud Barnard~68, at a distance of 500~ly$^1$  and 0.5~ly in diameter.(Image Credit: Wikipedia)}
\label{b68}
\vskip -12pt
\centering
\includegraphics[width=5.5cm, height=5.5cm]{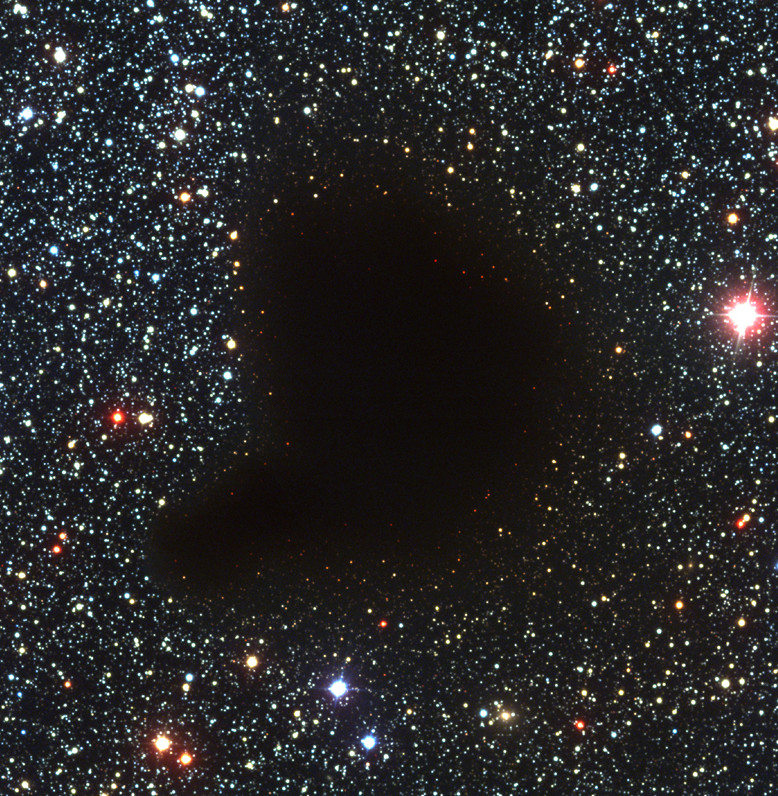}
\end{figure}

There are regions in the sky that appear as dark patches (Fig \ref{b68}\footnote{B = 445 nm, V = 551 nm, I = 806 nm}) due to extinction which is absorption and scattering caused by intervening matter. Extinction is inversely proportional to wavelength and hence observations in longer wavelengths are used.

 The star count method was used to estimate the missing number of stars in a region by  comparing them to the number of stars in the neighbouring region. These regions are identified as Giant Molecular Clouds (GMCs) and have sizes of  20-100 pc, masses ranging from $10^4-10^6$ M$_{\odot}$, $n \approx$ 50-100 cm$^{-3}$ and  $T \approx 10$ K. They are confined to the  Galactic plane, coinciding  with the  spiral arms and concentrated in a  ring 3.5-7.5 kiloparsec (kpc) in diameter.
 
\rightHighlight{Molecular clouds are the cold, dense regions where SF formation takes place. They are embedded in gas and dust and hence have to be observed at longer wavelengths from near infrared to radio. Young stars also emit very strongly in the Xray. Hence multiwavelength observations are required to get a complete census of young stars in GMCs.}‏
The mass of GMCs is mainly contained in molecular hydrogen and helium atoms. About one percent is made up of dust, typically silicates and/or graphites. Other molecules and their isotopes like  CO, NH$_3$, CN, H$_2$O have been detected, with CO being the most abundant.  Molecular hydrogen does not possess a permanent dipole moment. Hence, at the typical temperatures of molecular clouds it does not emit any radiation.
Observations use different tracers such as dust or other abundant  molecules and indirectly estimate the amount of molecular hydrogen. They can be observed in the radio by observations of molecular lines, or using dust extinction. 

 Understanding the formation and evolution of GMCs, is a daunting problem. One immediate issue is the vast range in scales between galaxies and protostellar discs, from ~10 kpc  to ~10$^{-3}$ pc. Another difficulty in the study of molecular clouds is the complex physics involved: gravity, magnetic fields, thermodynamics, turbulence and stellar feedback. The Interstellar Medium itself is a multiphase medium of atomic, molecular and ionized hydrogen  with a range of temperatures from 10 K to 10$^8$ K and large differences in density.

\section{Jeans instability}
\begin{figure}[!t]
\caption{Star Formation Image credit: http://science.howstuffworks.com/how-are-stars-formed.htm}
\label{sf}
\vskip -12pt
\centering
\includegraphics[width=7.5cm, height=7.5cm]{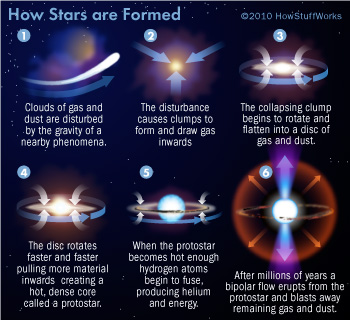}
\end{figure}

Sir James Jeans proposed a very simple theory for the formation of stars, based on the Kant-Laplace nebular hypothesis. An interstellar cloud is in hydrostatic equilibrium, i.e., there is a balance between the gravitational force and the gas pressure. 
\leftHighlight{James Jeans proposed gravitational collapse as the mechanism for formation of stars. The conditions for this to take place was that the GMC is cold and dense. As the core gains mass, material accretes on to it, forming a disc. When the temperature of the core rises to 10$^6$ K, nuclear fusion begins and a star is born.}
He proposed that if a cloud was cold and dense enough, then the gravitational force would dominate, leading to the gravitational collapse of the cloud (Fig \ref{sf}). As the cloud collapses it breaks into fragments in a hierarchical fashion, until the fragments  are close to stellar mass.  Jeans Mass ($M_J$) is the characteristic mass of a cloud when this condition gets satisfied and is given by:
$ M_J=3\times 10^{4}\sqrt{{\frac {T^{3}}{n}}} M_{\odot}$ where $T$ is the temperature and $n$ is the density.
The centre of the clump collapses to a  dense, gravitationally stable core  known as a protostar, which  heats up as it continues to contract. The protostar grows by accreting more material from the surrounding molecular cloud, and thus its core gets denser and hotter.
The protostar gravitates  material from the surrounding molecular cloud, which accretes onto the star. Due to the conservation of angular momentum,  this material spirals in towards the star and forms a disc of material that orbits the star, slowly accreting onto the star in bright bursts that illuminate the surrounding cloud. With each burst of accretion the star becomes hotter and more massivem till it  hot enough for nuclear fusion to take place. At first the star can only burn deuterium, but as it gets hotter it  burns hydrogen just like our own Sun. The star is now beginning to shine quite brightly and the radiation from the star prevents further material accreting onto the star and may even begin to disperse the remaining material in the disc that still surrounds the star.

\rightHighlight{YSOs are classified based on their spectral energy distribution. Class 0/I are the youngest which evolve to Class II and then the discless Class III sources. The spatial distribution of YSOs can be used to study the progress of SF formation and the influence of its environment. Images from ALMA of HL Tau should a very clear agreement of this picture of SF.}  

Once the  star has started nuclear fusion from  hydrogen into helium we say that it is born (Shu et al 1987). Hydrogen fusion is the natural source of energy for most stars  and the star continues fusion of lighter to heavier nuclei till it reaches the atomic number of iron. For nuclei heavier than iron, repulsive forces are stronger and fusion becomes an endothermic process, requiring energy. Elements heavier than iron are only produced in highly energetic events like a supernova explosion.
 
Young Stellar Objects (YSOs) are usually classified using criteria based on the slope of their spectral energy distribution, introduced by Lada (1987). He proposed three classes (I, II and III), based on the values of intervals of spectral index $ \alpha$ given by  $\alpha ={\frac {d\log(\lambda F_{\lambda })}{d\log(\lambda )}} $, where  $\lambda$, is the wavelength, and $F_{\lambda }$ is the flux density.

$\alpha$ is calculated in the wavelength interval of 2.2--20 $\mu$m. Later, Class 0 objects were added with strong submillimeter emission, but very faint at $\lambda <10 \mu$m and then, the class of `flat spectrum' sources were added.


This classification schema follows an evolutionary sequence, where the most deeply embedded Class 0 sources evolve towards Class I stage, dissipating their circumstellar envelopes. Eventually they become optically visible on the stellar birthline as pre-main-sequence stars.

\begin{figure}[!t]
\caption{ALMA image of the young star HL Tau and its protoplanetary disc. Credit: ALMA (NRAO/ESO/NAOJ); C. Brogan, B. Saxton (NRAO/AUI/NSF)}
\label{hltau}
\vskip -12pt
\centering
\includegraphics[width=4.5cm, height=3.5cm]{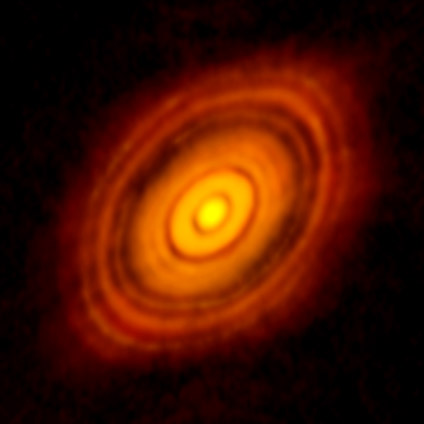}
\end{figure}

The most important result of this picture of SF is the co-formation of stars and planets from the protoplanetary discs and it has been recently validated by observational evidence by  the Atacama Large Millimeter/submillimeter Array (ALMA). 
Figure \ref{hltau} reveals in astonishing detail the planet-forming disc surrounding HL Tau, a Sun-like star located approximately 450 ly from Earth in the constellation Taurus.

\section{Filaments}
\rightHighlight{Star clusters are found where filaments overlap. The obvious question is whether the filament collision occurs before, after, or even during the SF process. Observational evidence supports all above scenarios and that could be possible because of the wide range of environments present, leading to realizations of all possible cases.} 

\begin{figure}[!t]
\caption{A far-infrared image of the Taurus molecular cloud, showing the filamentary structure of the gas
(Credit: Herschel Space Observatory)}
\label{fila}
\vskip -12pt
\centering
\includegraphics[width=5.5cm, height=3.5cm]{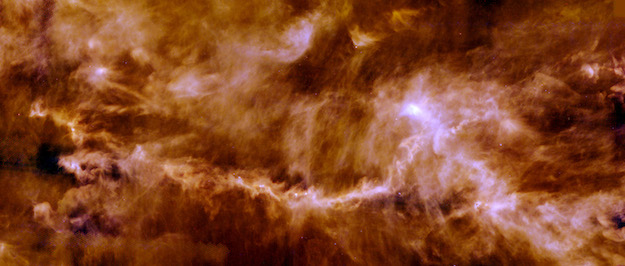}
\end{figure}

Molecular clouds have considerable structure, which is quite filamentary, formed by variable densities of gas and dust, causing variable extinction, very clearly seen in Herschel images (Fig. \ref{fila}).

This structure is thought to arise due to a combination of shock compression (due to collisions between material) and self-gravity (filaments can form gravitationally-stable structures on their own). These filaments can be seen on all spatial scales,  large and small (Andre et al 2014).

Many of these filaments are dense, containing many times the mass of our Sun in molecular gas. This high density implies that they can be gravitationally unstable, which can lead them to collapse and potentially form stars. Simulations of filaments suggest that a single filament can actually lead to the formation of multiple stars.
\begin{figure}[!t]
\caption{Star clusters forming in the Rosette Molecular Cloud
(Credit: Schneider et al. 2012)}
\label{sch}
\vskip -12pt
\centering
\includegraphics[width=7.5cm, height=8.5cm]{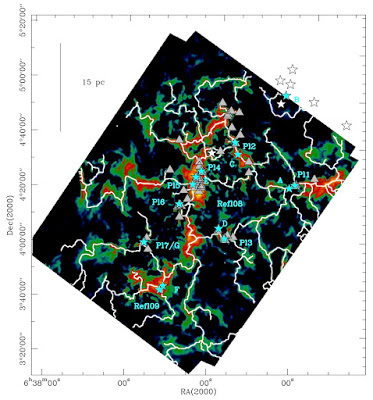}
\end{figure}

This picture where SF occurs in dense gas is seen in all the star forming regions and the youngest stars appear on the densest parts of the filaments, and on small scales (like NGC 1333) and on  much larger scales of giant molecular clouds such as the Orion A cloud.

Figure \ref{sch}  shows an image of filaments and star clusters in a star forming region known as the Rosette Molecular Cloud. The background image shows the distribution of dense gas in the cloud, with the density of the gas ranging from low-density (black) to high-density (green and red). Also are marked (in white) the positions of the filaments that make up the molecular cloud, and on top of that (the turquoise stars) are the positions of known star clusters, typically at points where filaments overlap. In somewhat older regions we start to see clusters of stars where there is no gas and the filamentary structure has dissolved, but the gas morphology could imply it being blown away by the young stars.

\section{Embedded Clusters}
\begin{figure}[h]
\centering
\includegraphics[width=7cm,height=6cm]{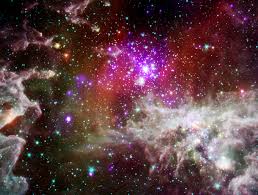}
\caption{This composite image of NGC 281 contains X-ray data from Chandra, in purple, with infrared observations from Spitzer, in red, green, blue.X-ray: NASA/CXC/CfA/S.Wolk; IR: NASA/JPL/CfA/S.Wolk}
\label{ngc281}
\end{figure}
Once stars form, embedded clusters are the earliest units of SF, of young stars  embedded in gas and  dust, invisible in the optical. They have sizes of 0.3-1 pc, and masses 20-1000 M$_{\odot}$, mean stellar  densities $1-10^3$ M$_{\odot}$ pc$^{-3}$. The SF Efficiency (SFE) is the ratio of the mass of stars to the total mass of the original cloud and is  10-30\%. It is this  unused gas that provides the gravitational glue that binds the cluster. Most clusters lose their gas in the first 5 Myr which leads to the disruption of clusters called  `infant mortality'. Barely  4 \% of  clusters survive beyond 100 Myr.

It is very important to study these objects in the early phase of their formation to understand their distribution, spatial as well as temporal, stellar, planetary companions, environments, etc.  

Embedded Clusters can be identified by making systematic searches of molecular clouds in infrared bands, say K  band (2.2 $\mu$m). Often, most members are obscured and the densities may not be significantly higher than the background. Other methods are surveys of signposts of SF like outflows, luminous IRAS sources, Herbig AeBe stars, etc.  We also use surveys using all sky data of 2MASS (Skrutskie et al, 2006), DENIS which can give good methods of statistical subtraction of stars in cluster and field areas to map over densities. 

\leftHighlight{Embedded clusters are the earliest signs of SF. The crucial step is detection of the YSOs because of the high levels of extinction}

For young embedded clusters with ages $< $ 3 Myr, at least half of the members will have circumstellar discs (Haisch et al 2001). Observations using the Spitzer Space Telescope at wavelengths 3 - 70 $\mu$m have been very useful to study  discs around stars (Winston et al 2007). For discless stars, a very effective method is  observations in the Xray, as young stars emit Xrays at levels $10^2-10^4$ times that of normal stars, particularly during the first 10 Myr of their lives (Winston et al 2007) (see Fig \ref{ngc281}).

\section{Some important aspects of Star Formation}

\subsection{Initial Mass Function (IMF)}
 The distribution of mass amongst the stars born from the same parent cloud is called the IMF. The IMF is often stated in terms of a series of power laws, where $N(M)$ the number of stars of mass $M$ within a specified volume of space is proportional to $M^{- \alpha}$ where $\alpha$ is a dimensionless exponent. The form of the IMF for stars with $ M > 1 M_{\odot}$ was discovered by Salpeter (1955), who found  $\alpha = 2.35$.

\subsection{Multiplicity}
\begin{figure}[h]
\centering
\includegraphics[width=8cm,height=6cm]{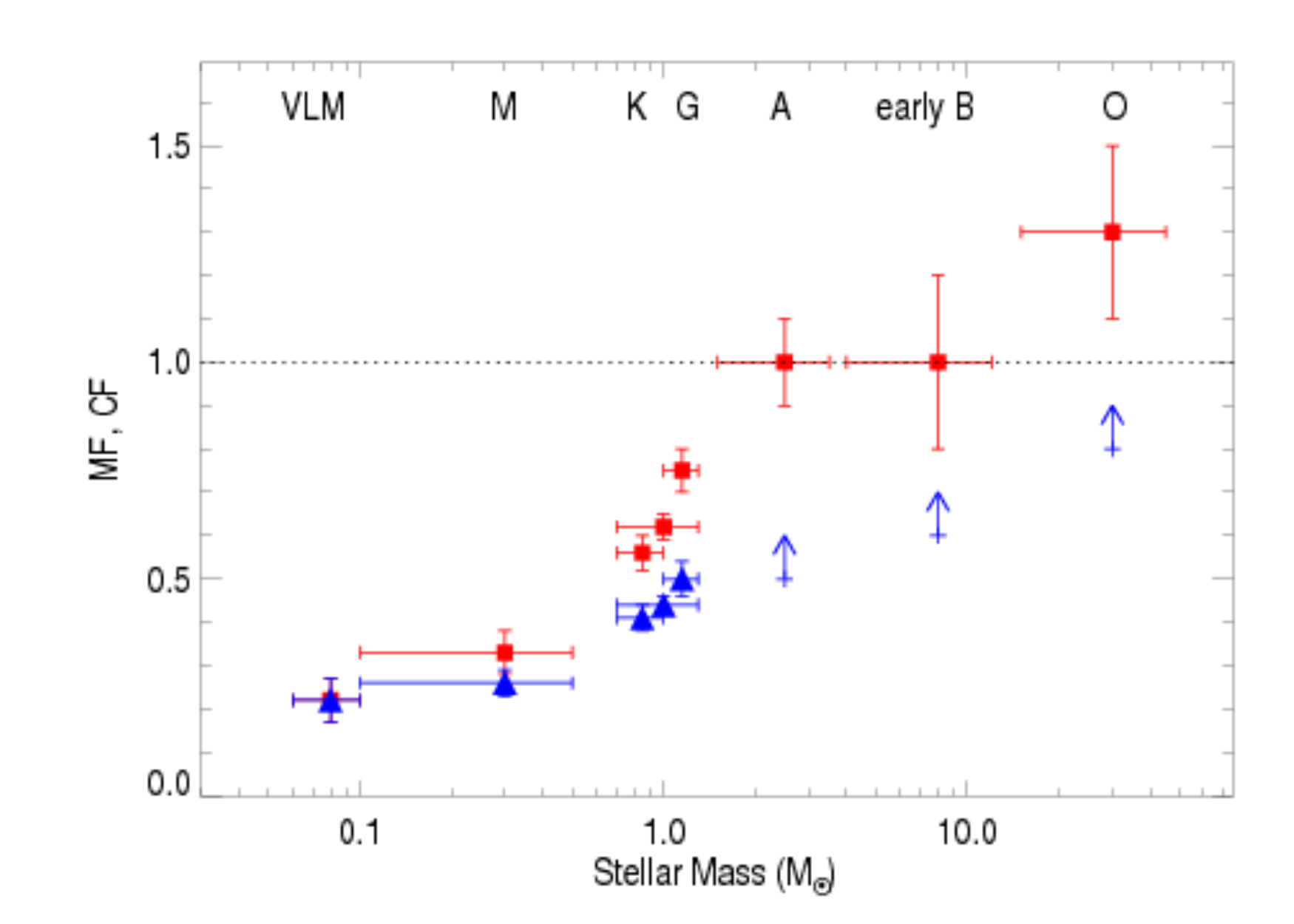}
\caption{Dependency of CF (red squares) and MF (blue triangles) with primary mass for MS stars (Duchene and Krauss, 2013).}
\label{mul}
\end{figure}

It was earlier believed that most stars are born in multiples (Larson 1972). It has been also found that there is a linear relation between the mass of a star and the number of multiples. Figure~\ref{mul} shows the relation between the frequency of multiple systems and the companion frequency  MF and CF, resp, to the mass of the star. For M type stars 66-67\% stars are single.  Recent studies of the IMF, show that the IMF breaks from a single power law near 0.5M$_{\odot}$ and has a broad peak between 0.1--0.5M$_{\odot}$ (Muench et al 2002). On either side of this peak the IMF falls rapidly. The peak of the IMF lies around M type stars  and it has been estimated  that 73\% -78\% of all stars are M type. Therefore, two-thirds  of (main sequence) stars currently residing in the galactic disc are single stars (Lada  2006). However, it is found that there is a decline in multiplicity as we go from Class 0 to Class I, II and III stars, implying that all single stars may not have been born single.

The questions that follow are: Is this how stars are formed? Individually? Or do stars form in binaries and multiples which later disrupt? Then, how do we explain the dependence of stellar multiplicity with mass? Or does increased cloud turbulence in massive dense cores lead to efficient core fragmentation and higher incidence of binary stars?

\subsection{Mass Segregation}
Mass segregation is the spatial distribution of stars according to their masses. It is observed that there is a concentration of high-mass stars near the centre and the low-mass ones away from the centre. This can take place as a result of dynamical interactions between stars in the clusters or could be primordial. The variation of the MF of clusters in different regions of the clusters has been studied by (Hasan and Hasan 2011 and references therein). 


\section{Conclusion}
\rightHighlight{Many unsolved problems of this process continue to puzzle astronomers like: How and why are stars clustered when they form? What causes stars to form with different masses? Is there a different process of SF for low and high mass stars? What brings the SF process within a molecular cloud to a halt? How do stars form in diverse environments? How do massive stars influence the formation of low mass stars? How coeval is SF?}
Stars are forming in our galaxy at a rate of between 1 and 4 $M_{\odot}$yr$^{-1}$. In contrast to elliptical galaxies, star formation is still going on in spiral galaxies because of their reservoirs of molecular gas which is the fuel for new stars.

 The optically dark molecular clouds are nurseries of stars, where the enigmatic process of SF takes places under the grip of gravity.  New observations and further research is required in these areas to answer some long-standing questions about the universe in which we live and to decipher the secrets in the starlight in darkness.
 


\section*{Acknowledgement}
The author would like to thank the referee for her/his valuable comments that helped improve the content of the article.  The author also thanks Prajval Shastri, Rajaram Nityananda and colleagues who came up with the lovely idea of this issue of Resonance - indeed a great way to commemorate the contribution of Women in Science!
%


\end{document}